\documentclass[aps,pra,superscriptaddress,twocolumn,10pt]{revtex4-2}
\usepackage{bm}
\usepackage{amsmath}
\usepackage{amsfonts}
\usepackage{amssymb}
\usepackage{graphicx}
\usepackage{color}
\usepackage[colorlinks,linkcolor=blue,anchorcolor=blue,citecolor=blue,urlcolor=blue]{hyperref}
\usepackage{dcolumn}

\begin{document}

\title{Thouless pumping and trapping of two-component gap solitons}
	
\author{Hao~Lyu}
\email{hao.lyu@oist.jp}
\affiliation{Quantum Systems Unit, Okinawa Institute of Science and Technology Graduate University, Onna, Okinawa 904-0495, Japan}

\author{Yongping~Zhang}
\email{yongping11@t.shu.edu.cn}
\affiliation{Institute for Quantum Science and Technology and Department of Physics, Shanghai University, Shanghai 200444, China}
	
\author{Thomas~Busch}
\email{thomas.busch@oist.jp}
\affiliation{Quantum Systems Unit, Okinawa Institute of Science and Technology Graduate University, Onna, Okinawa 904-0495, Japan}

\begin{abstract}

We study Thouless pumping and arresting of gap solitons in a two-component Bose gas loaded into an optical superlattice. We show that, depending on the atomic interactions and chemical potentials,
the two solitons can be simultaneously pumped or trapped, but we also identify regimes where one soliton is pumped and the other arrested. These behaviors can be understood by considering an effective model of the system based on a variational approach, which reveals that the
soliton width is a suitable qualifier for the observed behaviors:
solitons with a larger width get transported, while solitons with a smaller width get trapped. Since these width can be controlled by tuning interactions, it should be possible to observe all behaviors experimentally.

\end{abstract}
	
\maketitle

{\it Introduction}. The concept of Thouless charge pumping uses a periodically driven one-dimensional lattice to realize quantized motion of particles under adiabatic conditions~\cite{Thouless1,Thouless2,Citro}. 
This process is of topological nature, which means that the particle motion can be characterized by the Chern number.
In the recent years,  Thouless pumping of particles has been experimentally realized in a variety of systems, 
such as quantum dots, ultracold atoms, optical waveguides, and acoustic materials~\cite{Switkes,Lohse,Nakajima,Schweizer,Minguzzi,Kraus,Zilberberg,Xia,LiJ}, and it
provides a practical route for realizing directional transport of particles and exploring high-dimensional topological physics~\cite{Nakajima2021,Lohse2022}.

Interactions between particles are known to change energy bands in topological lattices, which can lead to a variety of novel phenomena~\cite{Cooper}.
The impact of interaction effects on Thouless pumping is therefore interesting and has been studied intensely in recent years.
Notably, if the interactions are weak enough to keep the energy gap open, 
the displacement of particles can still be characterized by the Chern number of the linear system~\cite{Tangpanitanon,Haug,Luo,Wawer,Kuno2021,Wu,ZhuZ}. 
Stronger interactions, on the other hand, can lead to more complicated particle transport phenomena and can even induce topological phase transitions. 
When the interactions lead to the energy gap closing, the system has a zero Chern number, 
and Thouless pumping is broken~\cite{Qin,Nakagawa2018,Hayward,Walter}.
On the other hand, interaction-induced particle pumping can be realized if the interaction leads to a nonzero Chern number
~\cite{Nakagawa,Stenzel2019,vanVoorden,Stenzel2020,Kuno2020,Gonzalez-Cuadra,Viebahn}.
Furthermore, multiple quasidegenerate ground states can be induced by long-range interactions so that fractional charge pumping can be achieved~\cite{Zeng,Taddia}.

In interacting systems solitons are of fundamental interest, as they represent self-trapped states based  
on a balance between the dispersion and a nonlinearity~\cite{Lederer,Kartashov}.
Nonlinear periodic lattices support gap solitons, which are solutions whose chemical potential lies in the energy gap of the linear spectrum~\cite{Eggleton,Eiermann,Pan}.
The existence of gap solitons is therefore highly sensitive to the energy band structure and the nonlinearity strength and it is therefore hard to manipulate and transport gap solitons in periodic lattices.
However, the mechanism of Thouless pumping has recently been explored as a robust way to study transport in  such systems~\cite{Jurgensen2021,Jurgensen2022,Mostaan,Fu,Fu2D,Jurgensen2023} and for weak enough interactions that do not close the energy gap, it is possible to realize Thouless pumping of solitons by periodically driving the lattice.
In fact, integer and fractional displacements of optical solitons have already been observed in optical waveguide arrays~\cite{Jurgensen2021,Jurgensen2023}.

Ultracold atoms provide a versatile platform for studying soliton physics due to their highly controllable atomic interactions and engineerable dispersion relations~\cite{Kevrekidis,Mitchell,Sanz,Zhang:15}.
Gap solitons have been observed in Bose-Einstein condensates (BECs) loaded into optical lattices by pushing the wave packet into the energy gap via 
tuning of the lattice laser frequency~\cite{Eiermann}.
Furthermore, a two-component gas can provide more degrees of freedom and different types of solitons have been revealed by tuning intra- and interspecies interactions~\cite{Gubeskys,Adhikari,Farolfi,Bresolin}.
For example, the two solitons can exist in different energy gaps due to having different chemical potentials or competition between the intra- and interspecies interactions, 
which can lead to different transport properties. 
The transport fidelities of the two solitons may also strongly depend on the interspecies interactions.
In a recent work, a novel phenomenon of quantized pumping of impurities induced by atomic interactions in Bose mixtures has been studied~\cite{Mostaan}.
Here, we aim to clarify how different regimes of atomic interactions affect the transport or trapping of two-component gap solitons.

In this work, we therefore study Thouless pumping and trapping of gap solitons in a binary BEC loaded into an optical superlattice.
By tuning the atomic interactions and the chemical potentials, the two solitons can be transported or trapped, either together or individually.
To understand these phenomena, we use a Gaussian wave to approximate the soliton state and propose an effective model derived from the variational method. 
In particular, the effective model indicates that the soliton transport requires a soliton width above a critical value, which has not been revealed in previous works.
These results provides new insights in understanding Thouless pumping of solitons, which may be used to manipulate gap solitons in periodic potentials.

{\it Model}. 
We consider a binary BEC in an effectively one-dimensional geometry, which can be realized by tightly confining the motion of the atoms along two transversal directions.
In the longitudinal $x$ direction two lattice potentials of different wavelength (long and short) are applied, so that a superlattice structure is created.
We denote the wave vectors and lattice depths of the short and long lattices as $k_{s,l}$ and $V_{s,l}$, respectively, and assume that the frequency of the long lattice is periodically tuned to dynamically  modulate the superlattice.
We also assume that the lattice potentials affect both species identically, so that the superpotential can be written as
\begin{align}
V(x,t)=V_s\cos^2(k_sx) + V_l\cos^2(k_lx-\omega t),
\end{align}
with $\omega=\pi/T$ and $T$ being the driving period.
To use dimensionless physical quantities in all the equations, we scale the units of length, momentum, energy, and time as $1/k_0$, $k_0$, $\hbar^2k^2_0/m$, and $m/(\hbar k^2_0)$, respectively. 
Here $k_0$ is a quantity related to the wave vectors of the lattice lasers and $m$ is the atomic mass. 
For simplicity, we consider that the lattice period of the long lattice is twice that of the short lattice, $k_s/k_0=2k_l/k_0=2$.

The coupled Gross-Pitaevskii equations (GPEs) describing the above system can be written as
\begin{align}
i\frac{\partial \psi_1}{\partial t}=\left[ -\frac{1}{2}\frac{\partial^2}{\partial x^2} + V(x,t)  + g_{11}|\psi_1|^2 + g_{12}|\psi_2|^2\right] \psi_1, \notag\\
i\frac{\partial \psi_2}{\partial t}=\left[ -\frac{1}{2}\frac{\partial^2}{\partial x^2} + V(x,t)  + g_{22}|\psi_2|^2 +g_{12}|\psi_1|^2\right] \psi_2,
\label{eq:GPE}
\end{align} 
where $\psi_{1,2}$ are the wave functions of the two atomic species. The intra- and interspecies interaction coefficients are given by 
$g_{ij}$ ($i,j=1,2$) and are proportional to the corresponding $s$-wave scattering lengths.

In nonlinear lattice settings like the above, fundamental gap solitons correspond to nonlinear Bloch waves in the semi-infinite or first energy gap. 
Since the shape of these solitons can be approximated well by a Gaussian function~\cite{Zhang2009}, it is possible to apply a variational approach to study their dynamics~\cite{Trombettoni,Navarro}.
For this, the Lagrangian density of the system has the form
\begin{align}
\mathcal{L}=\sum_{j=1,2}\Bigg[ &
\frac{i}{2}\left( \psi^{\ast}_j\frac{\partial\psi_j}{\partial t} - \frac{\partial\psi^{\ast}_j}{\partial t}\psi_j\right)+\frac{1}{2}\left| \frac{\partial \psi_j}{\partial x} \right|^2 \nonumber\\  
&- V(x,t) |\psi_j|^2-\frac{g_{jj}}{2}|\psi_j|^4\Bigg] -g_{12}|\psi_1|^2 |\psi_2|^2,
\label{eq:Lagranian}
\end{align}
and an ansatz for each soliton solution can be constructed as
\begin{align}
\psi_{j}=A_j\exp\left[-\frac{(x-x_j)^2}{2\sigma^2_j}+ia_jx^2+ib_jx+ic_j\right].
\label{eq:gaussian}
\end{align}
Here, $A_j$, $x_j$ and $\sigma_j$ characterize the amplitude, center position, and width of the Gaussian wavepacket, respectively, while
$a_j$, $b_j$, and $c_j$ are coefficients related to the phase of the wave function. 
All these coefficients are time-dependent, due to the dynamic modulation of the superlattice.
Note that a similar approach was previously used to analyze mobile solitons in optical lattices driven by the gravity~\cite{Trombettoni}.

\begin{figure}[ht]
\includegraphics[width=3.4in]{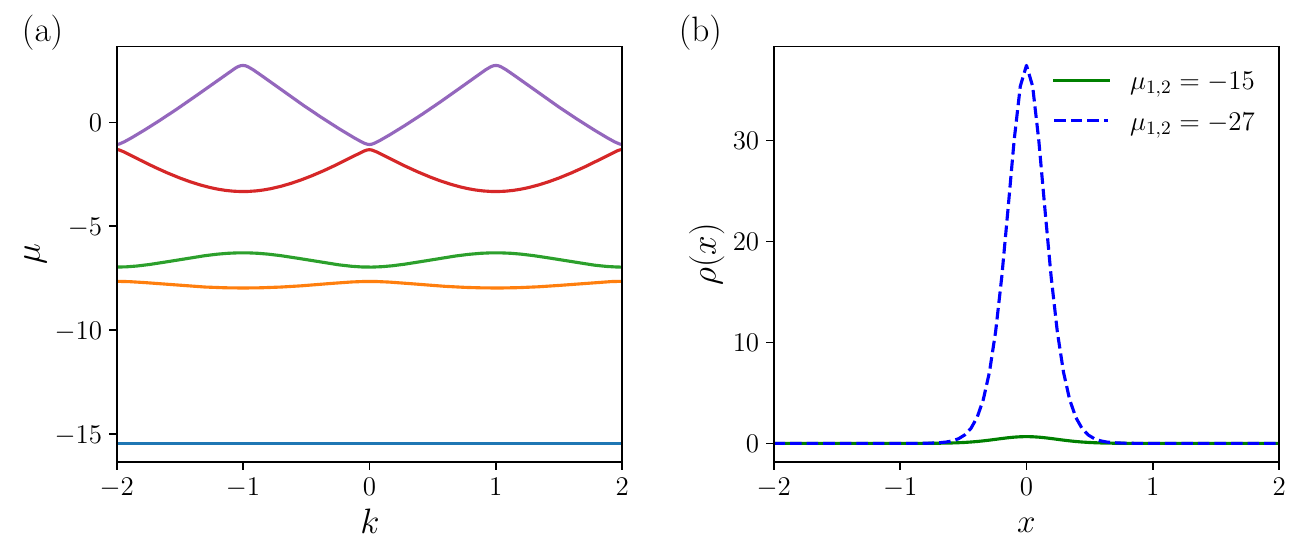}
\caption{Gap solitons in optical superlattices. (a) Spectrum of the linear system when the lattice depths are chosen as $V_s=V_l=-10$. (b) Total density profiles of two gap solitons for different parameters.
The green solid line shows a soliton in the first energy gap ($\mu_{1,2}=-15$) with $g_{11}=g_{22}=g_{12}=1$, and the blue dashed line corresponds to a soliton in the semi-infinite gap ($\mu_{1,2}=-27$) with $g_{11}=g_{22}=1$ and $g_{12}=-2$.
}
\label{fig1}
\end{figure}

Substituting the Gaussian ansatz into Eq.~(\ref{eq:Lagranian}) and integrating the resultant Lagrangian density in position space, gives 
\begin{align}
L&=\sum_{j=1,2}\sqrt{\pi}A^2_j \sigma_j \left[ 
f_j-(2a_jx_j+b_j)^2 + \frac{g_{jj}A^2_{j}}{2\sqrt{2}}   \right. \notag\\
&\phantom{={}}\left. -\frac{ V_se^{-k^2_s\sigma^2_j}}{2} 
\cos(2k_sx_j)-\frac{ V_{l}e^{-k^2_l\sigma^2_j}}{2}\cos\left(2k_lx_j-2\omega t\right) \right] \notag\\
&\phantom{={}}-\frac{\sqrt{\pi}\sigma_1\sigma_2}{s}g_{12} A^2_1A^2_2
e^{-\frac{(x_1-x_2)^2}{s^2}}, 
\end{align}
where
\begin{align}
f_j&=-\frac{1}{2}\dot{a}_j\sigma^2_j-\dot{a}_jx^2_j-\dot{b}_jx_j-\dot{c}_j -\frac{1}{2\sigma^2_j}-2a^2_j\sigma^2_j, \notag\\
s&=\sqrt{\sigma^2_1+\sigma^2_2}.\notag
\end{align}
We can then obtain the equations of motion for the six variational coefficients by using the Euler-Lagrange equation, and since we are interested in  transport, we focus on the displacement of the center position of the Gaussian waves. This leads to
\begin{align}
\ddot{x}_1&=2k_sV_se^{-k^2_s\sigma^2_1}\sin(2k_sx_1)+2V_{l}e^{-k^2_l\sigma^2_1}\sin\left(2k_lx_1-2\omega t \right)  \notag\\
&\phantom{={}} +\frac{4\sigma_2}{s^3} g_{12}A^2_2(x_1-x_2)e^{-\frac{(x_1-x_2)^2}{s^2}}, \label{eq:motionx1}\\
\ddot{x}_2&=2k_lV_se^{-k^2_l\sigma^2_2}\sin(2k_sx_2)+2V_{l}e^{-k^2_l\sigma^2_2}\sin\left(2k_lx_2-2 \omega t \right)  \notag\\
&\phantom{={}}+\frac{4\sigma_1}{s^3} g_{12}A^2_1(x_2-x_1)e^{-\frac{(x_1-x_2)^2}{s^2}}. 
\label{eq:motionx2}
\end{align}
It is worth noting that the above equations of motion are independent of the intra-species interactions $g_{jj}$, as one would expect.
Equations~\eqref{eq:motionx1} and \eqref{eq:motionx2} describe the motion of two particles in a time-dependent lattice,
and therefore establish an effective model for soliton dynamics.

\begin{figure*}[ht]
\includegraphics[width=6.6in]{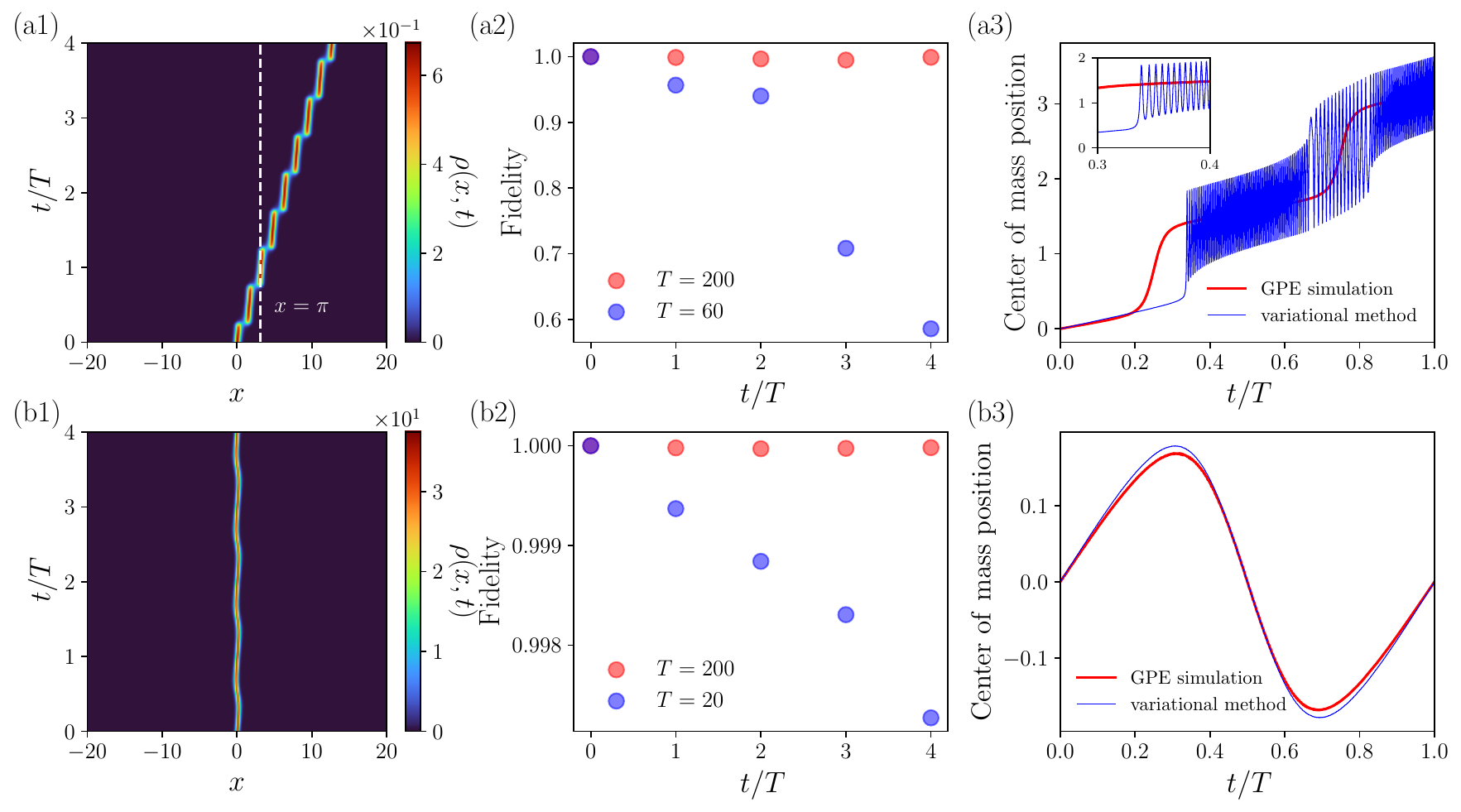}
\caption{Nonlinear Thouless pumping and trapping of two-component gap solitons.
(a1) Time evolution of the total density in the transport regime. (a2) The fidelity of the soliton wave function as defined by Eq.~\eqref{eq:fidelity} at integer values of $t/T$.
(a3) The center of mass position as a function of the time $t$,
where the red and blue lines represent results calculated from the GPEs and variational method, respectively.
The interactions are $g_{11}=g_{22}=g_{12}=1$.
(b1)-(b3) show the same quantities as in (a1)-(a3), but in the localization regime. 
The interactions are $g_{11}=g_{22}=1$ and $g_{12}=-2$. 
The width parameter in the variational method is $\sigma=0.5$ in (a3) and 0.3 in (b3).
The period of driving is set as $T=200$ in (a1), (a3), (b1) and (b3).
Other parameters are $V_{s}=V_{l}=-10$ in all the figures.}
\label{fig2}
\end{figure*}

{\it Simultaneous transport of two solitons}.
It was recently shown that solitons can be found in the semi-infinite gap in attractively interacting two-component systems in periodically driven superlattices. If the atom number is below a critical value, the solitons can be pumped across the lattice, otherwise they are localized~\cite{Fu}.
However, since for repulsive interactions solitons can also be found in the finite gaps, it is interesting to ask if these can also be pumped.

We will therefore focus on the pumping of two-component solitons with tunable inter-species interactions and chemical potentials, while
assuming that the intraspecies interactions are fixed and repulsive as $g_{11}=g_{22}=g=1$.
In this situation, the two soliton components have same density distributions and $g+g_{12}$ can be treated as an effective interaction. 
As a result, the system is similar to the single-component case,
and soliton pumping or trapping can be achieved by tuning the inter-species interaction and chemical potential.

To find the stationary soliton solutions of the coupled GPEs, we first calculate the spectrum of the linear system for the lattice configuration at $t=0$. For  deep lattices, i.e., $V_s=V_l=-10$, the spectrum possesses a broad first gap [see Fig.~\ref{fig1}(a)], 
in which soliton solutions can be easily found.
By substituting the wave function ansatz $\psi_j(x,t)=\psi_j(x)e^{-i\mu_jt}$ into Eqs.~\eqref{eq:GPE}, 
we obtain stationary GPEs, where the $\mu_j$ are the respective chemical potentials.
The stationary equations can then be solved numerically as a function of $g_{ij}$ or $\mu_j$ by using a Newton relaxation method. 
Two examples of solitonic density distributions in different energy gaps are shown in Fig.~\ref{fig1}(b), with 
the solid green line representing a solution in the first gap with $\mu_{1,2}=-15$ and for $g_{12}=1$.
For this chemical potential, the numerical calculations show that soliton solutions can always be found for $g_{12}\geqslant-0.9$. 
The dashed blue line in Fig.~\ref{fig1}(b) corresponds to a soliton in the semi-infinite gap with $\mu_{1,2}=-27$ for $g_{12}=-2$.
For this chemical potential, we find that the existence of soliton solutions requires $g_{12}\leqslant-1.2$.
As in this case the effective interaction, $g+g_{12}$, is negative,
the soliton state requires a large atom number to match the large negative chemical potential.
As a result, the soliton in the semi-infinite gap has a larger atom number than that in the first gap.

With the stationary soliton solutions at hand, we can then study the soliton dynamics by solving Eqs.~(\ref{eq:GPE}) numerically.
Starting with a soliton in the first energy gap with the same parameters as used for the state shown in Fig.~\ref{fig1}(b) (see the solid green line), we induce transport by modulating the superlattice on a slow time scale. The resulting time evolution of the total density is shown in Fig.~\ref{fig2}(a1).
After one period of driving, the lattice potential is the same as the initial one,
and the soliton density is displaced by $\pi/k_l=\pi$ [indicated by the white dashed line in Fig.~\ref{fig2}(a1)], which is a signature of Thouless pumping.
To characterize the transport, we introduce a fidelity related to the overlap of the evolving soliton state $\psi_j(x,t)$ and the respective translated state $\psi_{j}(x-\omega t/k_0,0)$ as 
\begin{equation}
\label{eq:fidelity}
F_j=\frac{\int dx \psi^\ast_j(x,t) \psi_j(x-\omega t/k_0,0)}{\int dx|\psi_j(x-\omega t/k_0,0)|^2}.
\end{equation}
This fidelity for times where $t/T$ is an integer is shown in Fig.~\ref{fig2}(a2).
One can see that the fidelity for slow evolutions ($T=200$, red dots) is very close to one even after successive pumping cycles, which indicates that the transport is very robust as expected from a topological process. 
However, one can also see that the fidelity declines for successive pumping steps when the adiabatic condition is not satisfied ($T=60$, blue dots). 
In this case, the soliton amplitude starts oscillating and excited modes are starting to be occupied. Additionally, 
the center-of-mass displacement oscillates as well.

To better understand the soliton transport let us employ an effective model obtained from the variational approach.
For $g_{11}=g_{22}$ and $\mu_1=\mu_2$, the two soliton components have the same wave functions and their dynamics is also the same.
Therefore, Eqs.~(\ref{eq:motionx1}) and \eqref{eq:motionx2} can be reduced to an equation of motion for an effective single particle as
\begin{align}
\ddot{x}_c&=2k_s V_s e^{-k^2_s\sigma^2} \sin(2k_sx_c) \notag\\
&\phantom{={}}+2k_lV_{l}e^{-k^2_l\sigma^2}\sin\left(2k_lx_c-2\omega t\right) ,
\label{eq:single}
\end{align}
where we have set $x_1=x_2=x_c$ and $\sigma_1=\sigma_2=\sigma$.
The first term on the right side of Eq.~\eqref{eq:single} represents the time-independent force induced by the short lattice,
while the second term is the time-periodic force resulting from the dynamical long lattice. 
We choose the initial condition for solving Eq.~\eqref{eq:single} as $x_c=\dot{x}_c=0$, so that the particle does not move in the absence of the driving force proportional to $V_l$.
One can see that in this effective model the motion of the particle is strongly related to the width of the Gaussian wave. 
For $k_s |V_s| e^{-k^2_s\sigma^2}<k_l|V_l|e^{-k^2_l\sigma^2}$, the time-dependent force dominates the particle motion, so that particle transport can be expected.
In contrast, the time-independent force dominates when $k_s |V_s| e^{-k^2_s\sigma^2}>k_l|V_l|e^{-k^2_l\sigma^2}$ and the soliton can be trapped.
By assuming $k_s/k_l,k_sV_s/(k_lV_l)>1$ and $V_{s,l}<0$, 
the critical width can be obtained by setting $k_sV_s e^{-\sigma k^2_s}=k_l V_l e^{-\sigma^2k^2_l}$ as 
\begin{align}
\sigma_{\mathrm{cr}}=\sqrt{\frac{1}{k^2_s-k^2_l}\log \frac{k_sV_s}{k_lV_l}}.
\label{eq:width}
\end{align}
It can therefore be predicted that the particle is transported for $\sigma>\sigma_{\text{cr}}$, while it is trapped for $\sigma<\sigma_{\text{cr}}$. 
We note that Eq.~(10) does not apply in the limits where solitons can become very extensive, such as shallow lattice settings.
In such a situation the critical width should also become large, which can not be predicted by Eq.~\eqref{eq:width} when $V_s/V_l$ is fixed.

To confirm this, we compare the results of the variational method with the ones obtained from solving the coupled GPEs numerically for a initial state of the same width in Fig.~\ref{fig2}(a3).
One can see that, after an initial delay, the center of the Gaussian wave with $\sigma=0.5$ is transported while oscillating rapidly (blue line) around the exact motion obtained by numerically solving the full GPEs (red line).
The rapid oscillations can be attributed to the effective potential $V_{\text{eff}}=V_se^{-\sigma^2k^2_s}\cos(2k_sx)+V_le^{-\sigma^2k^2_l}\cos(2k_lx-2\omega t)$, in which the two parts induce oscillation modes with frequencies $2k_{s,l}\sqrt{|V_{s,l}}|e^{-\sigma^2k^2_{s,l}/2}$, respectively. The combination of the two parts leads to complex dynamics of the displacement.
However, the effective model clearly captures the transport induced by the Thouless pumping.  

An example of arrested transport in the semi-infinite gap is shown in Fig.~\ref{fig2}(b1)-\ref{fig2}(b3) for a solution with the same parameters as those of the state indicated by the dashed blue line in Fig.~\ref{fig1}(b).
This localization phenomenon can also be understood within the effective model and
Fig.~\ref{fig2}(b3) shows the center of mass position of the atoms by solving the GPEs (red thick line) and the single-particle equation of motion (blue thin line). 
In this case the wave function has a narrower width ($\sigma=0.3$),
and therefore the time-independent force in Eq.~(\ref{eq:single}) dominates.
The two curves agree well, which indicates the validity of the effective model. 
In this case the fidelity of the wave function is high even for a fast driving [see the blue and red dots in Fig.~\ref{fig2}(b2)].

Having understood from the effective model that the soliton transport  requires the soliton width to exceed a critical value,
we have confirmed that the trapping phenomena only occurs for negative interactions which lead to a narrow soliton density distribution.
If all interactions are repulsive, the atom number of the soliton increases when the chemical potential comes close to the bottom of the second energy band.
However, the density distribution of the soliton is also broadened, so that the solition cannot be trapped. 
From the numerical calculations we also find that the critical width for a trapped Gaussian wave is about 0.49, which is even larger than the width of the transported soliton shown in Fig.~\ref{fig1}(b). 
Therefore, the effective model provides a qualitative analysis of soliton transport, and the critical soliton width cannot be predicted by Eq.~(\ref{eq:width}) exactly.
In addition, we note that the transported soliton has a small amplitude while the trapped soliton has a large one, which indicates that the amplitude should have a critical value above which the soliton is trapped.
Furthermore, the ratio of the atom numbers of the transported soliton and trapped one shown in Fig.~\ref{fig2} is about 0.03,
while the ratio of the soliton widths is $1.55$, which are both of the same orders of magnitude as the findings in Ref.~\cite{Fu}.

\begin{figure}[tb]
\includegraphics[width=\linewidth]{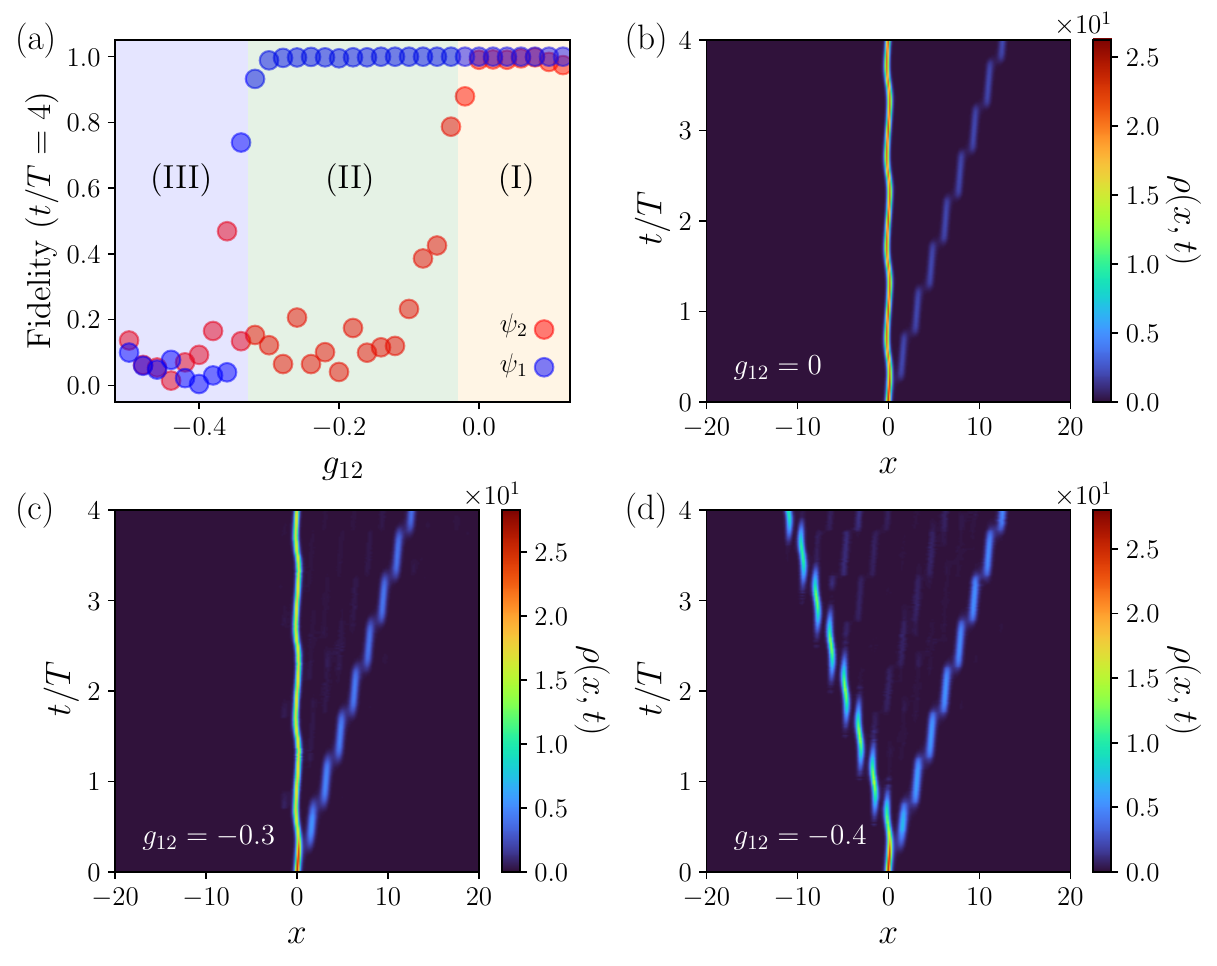}
\caption{
Thouless pumping and trapping of different soliton components with $g_{11}=1$ and $g_{22}=-1$.
(a) Fidelity of soliton wave functions at $t/T=4$ as a function of $g_{12}$.
In region (I), the first component with a small atom number is transported, while the second component with a large atom number is trapped, and the fidelities of both solitons are high.
In region (II), the first component is transported with a low fidelity, while the second component is trapped and is stable. 
In region (III), both the solitons are unstable.
Time evolution of the total atom densities for examples of the three regions are shown in (b)-(d),
where we choose $g_{12}=0,-0.3$ and $-0.4$, respectively.
Other parameters are same as in Fig.~\ref{fig2}.}
\label{fig3}
\end{figure}

{\it Pumping and localization of different soliton components}.
In the two-component quantum gases, the atomic interactions and chemical potentials of each components can be tuned independently, which gives rise to a rich tapestry of soliton states~\cite{Gubeskys,Adhikari}. One can therefore also expect different dynamical behaviors of the two soliton components in periodically modulated optical lattices by choosing appropriate parameters.

In this section we therefore explore the possibility of observing transport and trapping for the individual components of the soliton simultaneously in systems with intraspecies interaction strengths of $g_{11}=-g_{22}=1$.
In this situation, the two soliton components have different widths, and the interspecies interactions play an important role in the motions of the two solitons.
To demonstrate the effect of finite $g_{12}$, we plot the fidelity of the two solitons at $t/T=4$ as a function of $g_{12}$ in Fig.~\ref{fig3}(a).
One can immediately identify three regions corresponding to different soliton dynamics.
In region (I) ($-0.03\leqslant g_{12}\leqslant 0.13$),  the first component is transported while the second one is trapped, and the fidelities of both solitons are high ($F_{1,2}>0.9$).
We show the time evolution of the total density for $g_{12}=0$ in Fig.~\ref{fig3}(b), where the respective chemical potentials are $\mu_1=-14$ and $\mu_2=-30$.
It is worth noting that the atom number of the trapped soliton is much larger than that of the transported soliton [c.f.~Fig.~\ref{fig1}(b)] and the observed behavior of the two solitons agrees well with previously reported results~\cite{Fu}.
Note that the atom number of the first component decreases with increasing $g_{12}$, and the numerical solutions show that the soliton disappears for $g_{12}>0.13$. Increasing $g_{12}$ on the repulsive side has therefore a similar effect as decreasing $\mu_1$.

From Fig.~\ref{fig3}(a) one can see that the fidelity of the first component to be transported is always high in region (I).
However, as for $g_{12}<0$ the number of atoms in the transported soliton increases, the fidelity decreases with increasing negative interaction strength.
In contrast, the trapped component of the soliton can be seen to be almost unaffected by the inter-component interaction for $-0.33\leqslant g_{12}<-0.03$ [region (II) in Fig.~\ref{fig3}(a)].
This is due to the second component having a large atom number, and therefore the intraspecies interaction is always larger than the interspecies interaction.
A typical example of the time evolution of the total atom density in this case is shown in Fig.~\ref{fig3}(c).
In this case, the soliton shape becomes unstable with excited modes becoming occupied, which is similar to the $g_{11}=g_{22}$ case. The atom numbers of the transported and trapped soliton for the $g_{12}=-0.3$ case are about 4 and 8, respectively. Therefore $g_{12}|\psi_1|^2$ only has little impact on the motion of the second component. 
If $|g_{12}|$ is reduced further ($g_{12}<-0.33$), both solitons are unstable [see the region (III) in Fig.~\ref{fig3}(a)]. In this case the second component is broadened and has a lower particle number, so that it cannot be trapped.

We can again use the effective model to understand how $g_{12}$ affects the motion of the individual solitons.
Here, we consider small values of $|g_{12}|$, since for a large $|g_{12}|$ soliton solutions cannot be found in the full model.
From Eq.~(\ref{eq:motionx2}) one can see that a positive $g_{12}$ decreases the effective depth of the short lattice for the first component, while a negative $g_{12}$ has the opposite effect.
This means that the critical width for soliton transport is smaller in the $g_{12}>0$ region and the soliton can be pumped.
In contrast, a negative $g_{12}$ leads to a large particle number in the first component, so that the soliton becomes unstable and is unfavorable for transport. 
For the second component the situation is different. 
In region (II), the intraspecies interaction has only a weak effect on the depths of the lattices since the particle motion $x_2$ is very small, so that the second-component soliton is always trapped.
The effective model therefore again provides an intuitive physical picture for understanding Thouless pumping and trapping of gap solitons.

We note that the soliton width is related to the populations of the atoms in different energy bands. As pointed out in Ref.~\cite{Fu}, a transported soliton with a large width mostly populates the first band with Chern number 1, while the trapped one with a narrow width can occupy some bands with the sum of the respective Chern numbers as 0. 
In addition, the single particle in the effective model also has similar occupations of the topologically nontrival bands.
In this work, the soliton width provides an alternative way to understand the transport or trapping of solitons.

{\it Experimental considerations}.
Thouless pumping has been experimentally observed in both bosonic and fermion atoms~\cite{Lohse,Nakajima,Nakajima2021}.
In this work, we have considered a two-component BEC with tunable interactions loaded into an optical superlattice.
This can experimentally be achieved by using $^{39}\mathrm{K}$ atoms, in which the intraspecies and interspecies interactions can be tuned via a Feshbach resonance or coherent coupling~\cite{Sanz}.
We also note that the system discussed above can also be realized in optical waveguide systems~\cite{Jurgensen2021}. 
In this situation, $\psi_{1,2}$ represent two optical modes, 
and the time $t$ in all the equations should be replaced by the propagating coordinate $z$. In this case the self- and cross-phase modulations can be controlled by tuning light intensities and frequencies.

{\it Conclusion}.
We have studied the possibilities to Thouless pump or trap two-component matter-wave gap solitons in optical superlattices.
Depending on the different atomic interaction strengths and chemical potentials, the two solitons can be simultaneously transported or trapped, but regimes where one soliton is pumped and another one is trapped can also be realized.
These phenomena can be understood by using an effective model derived from a variational approach using a Gaussian ansatz, and we have shown  that the transport and trapping phenomena are related to the soliton widths.
The soliton in the repulsive interaction regime possesses a large width, which allows for pumping, whereas a negative interaction leads to narrow solitons which can be arrested.
Our results provide a new route to studying transport and generally manipulating multi-component gap solitons via Thouless pumping.

{\it Acknowledgment}. This work was supported by Okinawa Institute of Science and Technology Graduate University, and also supported by the
Japan-China Scientific Cooperation Program between JSPS and NSFC Under Grants No. 120227414 and No. 12211540002. Y.Z.~is supported by National Natural Science Foundation of China with Grants No. 12374247 and No. 11974235, and  Shanghai Municipal Science and Technology Major Project (Grant No. 2019SHZDZX01-ZX04).

\bibliography{NTP}

\end{document}